# Study of spin pumping in Co thin film vis-à-vis seed and capping layer using ferromagnetic resonance spectroscopy

Braj Bhusan Singh, Sukanta Kumar Jena, Subhankar Bedanta [*]

*Laboratory for Nanomagnetism and Magnetic Materials (LNMM), School of Physical Sciences, National Institute of Science Education and Research (NISER), HBNI, P.O.- Bhimpur Padanpur, Via – Jatni, Odisha, Pin - 752050, India*

**Abstract**: We investigated the dependence of the seed (Ta/Pt, Ta/Au) and capping (Pt/Ta, Au/Ta) layers on spin pumping effect in the ferromagnetic 3 nm thick Co thin film using ferromagnetic resonance spectroscopy. The data is fitted with Kittel's equation to evaluate damping constant and *g*-factor. A strong dependence of seed and capping layers on spin pumping has been discussed. The value of damping constant ($\alpha$) is found to be relatively large i.e. 0.0326±0.0008 for the Ta(3)/Pt(3)/Co(3)/Pt(3)/Ta(3) (nm) multilayer structure, while it is 0.0104±0.0003 for Ta(3)/Co(3)/Ta(3) (nm). Increase in $\alpha$ is observed due to Pt layer that works as a good sink for spins due to high spin orbit coupling. In addition, we measured the effective spin conductance $g^{\downarrow\uparrow} = 2.00 \pm 0.08 \times 10^{18}$ m$^{-2}$ for the tri-layer structure Pt(3)/Co(3)/Pt(3) (nm) as a result of the enhancement in $\alpha$ relative to its bulk value. We observed that the evaluated *g*-factor decreases as effective demagnetizing magnetic field increases in all the studied samples. The azimuthal dependence of magnetic resonance field and line width showed relatively high anisotropy in the tri-layer Ta(3)/Co(3)/Ta(3) (nm) structure.

PACS numbers: 75.76.+j, 75.78.-n, 75.70.Cn, 76.50.+g

[*]Corresponding author:
Electronic mail: sbedanta@niser.ac.in

# INTRODUCTION

The generation of pure spin current and its effect on the switching of the magnetization by spin transfer torque via spin orbital torque has been subject of vivid research in last one decade. Apart from basic research this subject has also implications in potential devices based on spintronics/spin-orbitronics [1–3], spin torque- oscillator [4,5], magnetic random memory devices [6,7], magnonics [6] etc. A fundamental understanding of the exchange interaction and spin-orbital interaction at the interfaces will increase speed, energy efficiency, miniaturization of size and its multi-functional utilization [8]. Pure spin current is produced by asymmetric scattering of the two electrons with opposite spin angular momentum in the presence of spin orbit coupling at the interface which is known as spin Hall effect [1,3,9,10]. This is usually investigated via ferromagnetic resonance (FMR) in which a microwave excites the spin in ferromagnetic (FM) layers. The spin excitation in the FM layer generates and propagates pure spin current into paramagnetic heavy metal (NM). This phenomenon is called as "spin pumping" and the spin current in to NM is given by

$$J_S \hat{s} = \frac{\hbar}{4\pi} g^{\downarrow\uparrow} \vec{m} \times \frac{d\vec{m}}{dt} \quad \ldots\ldots\ldots\ldots\ldots\ldots\ldots\ldots\ldots\ldots\ldots\ldots\ldots\ldots\ldots\ldots\ldots\ldots\ldots\ldots\ldots\ldots\ldots\ldots\ldots\ldots\ldots\ldots\ldots\ldots\ldots(1)$$

where $\vec{m}$ is the reduced magnetization ($\frac{\vec{M}}{M_S}$), $\hbar$ is the reduced Planck's constant, and $g^{\downarrow\uparrow}$ is the effective spin mixing conductance which is governed by the transmission of the spin current through FM/NM interfaces [11]. The understanding of magnetic dynamics dissipation in the material is very important to get sustainable spin current. Normally, magnetic dynamics dissipates the pure spin current through Gilbert damping effect, which is related to the electronic structure of the ordered material. However, other damping mechanisms e.g. in-homogeneities in the sample, two magnon scattering and interfaces might also play important role to enhance the damping effect [12–16]. The interfaces are very important for creating pure spin current and its dissipation. In addition, it affects the damping properties critically by various seed and capping layers. Recently, research has been focussed to understand the effect of heavy metal as seed and capping layers [17–22]. In a recent report Tokac *et al*. [21] have shown that the effective spin conductance mixing strongly depends on the interfaces of seed (Ta/Cu) and capping (Cu or Ir) layers. They observed large effective spin mixing conductance for Ir as a capping layer due to its large spin orbit coupling. It is clear that effective spin mixing conductance critically depends on the symmetry of the interfaces with respect to the ferromagnetic layer. In order to understand the effect of seed and capping layers on spin pumping dynamics we fabricated the multilayer structure *Seed layer/Co/Capping layer*. The seed layer and capping layers are the combination of heavy metals like Ta, Pt, and Au. We studied thin films having symmetric (seed and capping layers are same) and asymmetric (seed and capping layers are different) interfaces around the FM layer.

**EXPERIMENTAL DETAILS**

The samples are deposited at room temperature in a high vacuum system (Mantis Deposition Ltd., UK) having base pressure better than $5.0 \times 10^{-8}$ mbar. The schematic of the sample structure is shown in Fig. 1(a). The Ta, Cu, and Co layers are deposited by dc sputtering. While Pt and Au thin films are prepared by rf sputtering and electron beam evaporation, respectively. We deposited two sets of samples on Si(100) substrates having native oxide layer. In the first set of the samples, we deposited the multilayers Ta(3)/HM$^T$/Co(3)/HM$^B$/Ta(3) (in parenthesis thickness is in nm) in which HM$^T$ and HM$^B$ stands for heavy metal for top and bottom layers, respectively. Pt and Au are taken as heavy metals for HM$^T$ and HM$^B$ due to their high spin orbit coupling. To compare the results, we

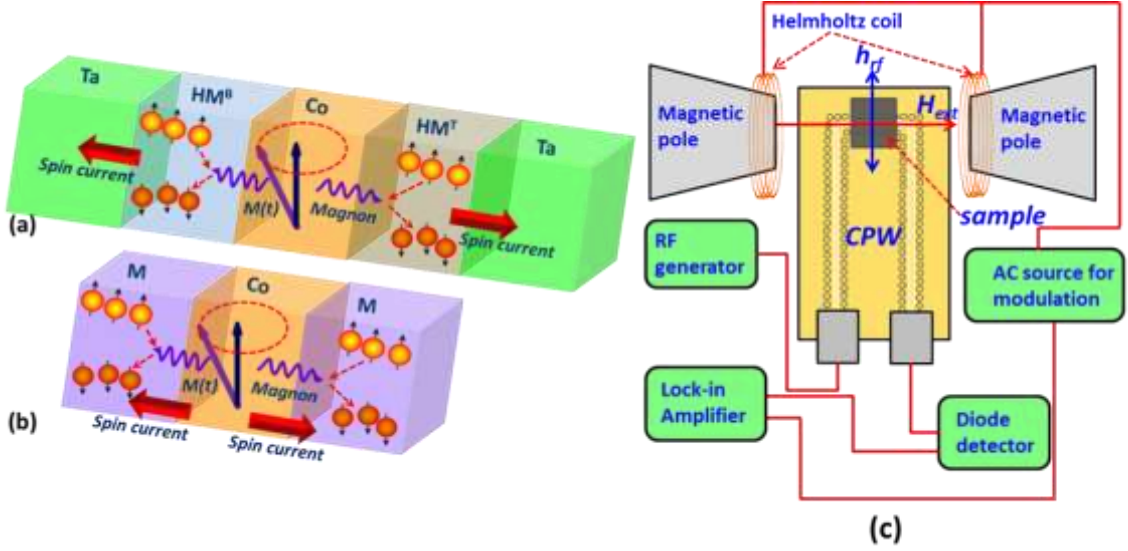

FIG 1(a) and (b) shows the schematics of generation of spin current in the multilayer and tri-layer sample structures (c) FMR measurement setup in which rf generator creates GHz frequency which works as a perturbation field ($h_{rf}$) perpendicular to external applied field ($H_{ext}$) and a diode detector through AC magnetic field modulation detects transmitted signal via lock-in based technique.

deposited second set of samples having tri-layer structure as shown in Fig. 1(b). The sample details are listed in Table I. The deposition rates for Ta, Cu, and Pt are kept to be same and are 0.018 nm/sec. Co is deposited at a rate of 0.016 nm/sec. The working pressure during sputtering deposition was $\leq 2.0 \times 10^{-3}$ mbar. A growth rate of 0.006 nm/sec is used to deposit Au thin film at $\leq 3.0 \times 10^{-7}$ mbar working pressure. The deposition rates are measured using quartz crystal monitor. FMR measurements are performed using NanoOsc Instrument Phase FMR in the frequency range of 5 to 17 GHz. The sample is kept in flip-chip manner on a 200 µm wide coplanar waveguide as shown in the Fig. 1(c). A lock-in amplifier based techniques is used to detect the signal in which rf field ($h_{rf}$) is perpendicular to the external magnetic field ($H_{ext}$).

Table I. The list and details of the fabricated samples.

| Sample Name | Sample details (In parenthesis thickness is in nm) |
|---|---|
| S1 | Si/Ta(3)/Pt(3)/Co(3)/Pt(3)/Ta(3) |
| S2 | Si/Ta(3)/Au(3)/Co(3)/Au(3)/Ta(3) |
| S3 | Si/Ta(3)/Au(3)/Co(3)/Pt(3)/Ta(3) |
| S4 | Si/Ta(3)/Pt(3)/Co(3)/Au(3)/Ta(3) |
| S5 | Si/Ta(3)/Co(3)/Ta(3) |
| S6 | Si/Cu(3)/Co(3)/Cu(3) |
| S7 | Si/Pt(3)/Co(3)/Pt(3) |

**RESULTS AND DISCUSSIONS**

Figure 2(a) shows the frequency dependent resonance field for the sample S1 to S4. The open symbols show the experimental data and the solid lines are the fits with Kittel resonance condition [23]:

$$f = \frac{\gamma}{2\pi} \sqrt{(H_K + H_{res})(H_K + H_{res} + 4\pi M_{eff})} \quad \ldots \ldots (2)$$

where $\gamma = \frac{g\mu_B}{\hbar}$ is gyromagnetic ratio; $H_K$, $H_{res}$, $g$, and $\mu_B$ are the in-plane anisotropy field, the resonance magnetic field, Lande $g$-factor and Bohr magneton, respectively, at the resonance frequency $f$. The effective demagnetizing field is given by

$$4\pi M_{eff} = 4\pi M_S + \frac{2K_S}{M_S t_{FM}} \quad \ldots \ldots (3)$$

where $M_S$ is the saturation magnetization, $K_S$ is the perpendicular surface anisotropy constant, and $t_{FM}$ is the thickness of the ferromagnetic layer.
The fitting with equation (2) to the frequency dependent resonance field graph gives the value of $g$ factor, effective demagnetization field ($4\pi M_{eff}$) and $H_K$. The data is well fitted with reduced chi-square ($\chi^2$) value ~ $10 \times 10^{-4}$. For fitting we used the procedure described by Shaw *et al.* [24]. The value of damping constant (α) has been evaluated using the equation [25]

$$\Delta H = \Delta H_0 + \frac{4\pi\alpha f}{\gamma} \quad \text{...............................................................................(4)}$$

where $\Delta H$, $\Delta H_0$, and $\gamma$ are the line width of the FMR spectra at frequency $f$, the inhomogeneous line width broadening and gyromagnetic ratio, respectively. The *g*-factor has been calculated from the gyromagnetic ratio value. It is known that the value of $\Delta H_0$ depends on the quality of the thin film [25,26].

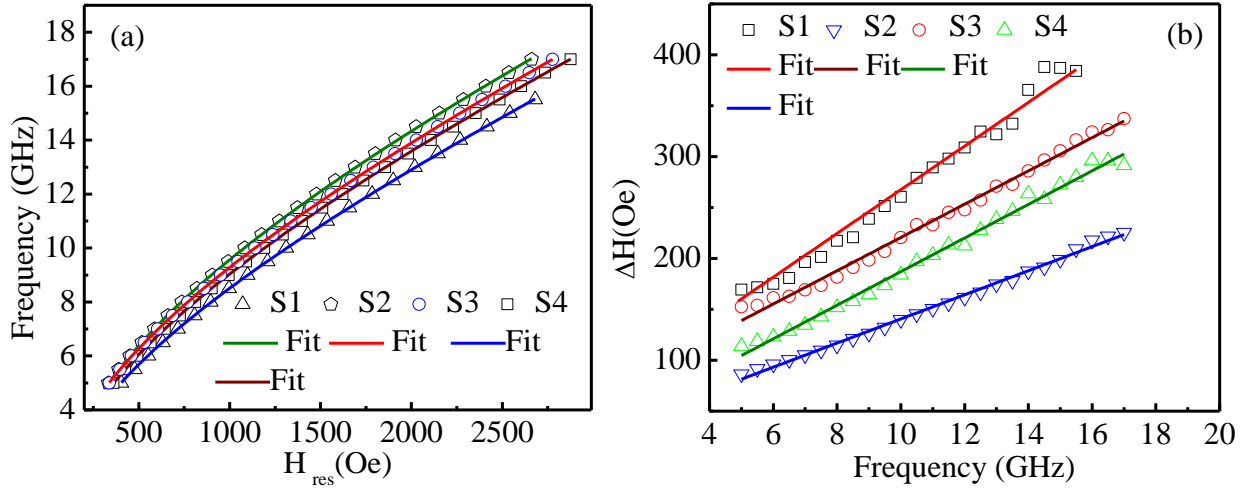

FIG 2(a) Resonance frequencies versus resonance magnetic field ($H_{res}$) for sample S1 to S4. Open symbols represent the experimental values while solid lines are fitted with equation (2). (b) Line width ($\Delta H$) versus resonance frequencies for sample S1 to S4. Open symbols are experimental data and solid lines are fitted with equation (4).

The fitted parameters $H_K$, $4\pi M_{eff}$, $\Delta H_0$, *g-factor,* and damping constant *(α)* of samples S1 to S4 are shown in the Table II. The sample S1 and S4 has the same seed layer (Ta/Pt) but different capping layers (Pt/Ta or Au/Ta) which make the multilayers structure with symmetric and antisymmetric capping layers. The extracted value of $4\pi M_{eff}$ is higher for the sample S4 in comparison to sample S1.

Table II: The parameters evaluated from the fitting of experimental data of sample S1, S2, S3 and S4 using equation (2) and (4).

| Sample | $H_K$ (Oe) | $4\pi M_{eff}$ (Oe) | g -factor | α | $\Delta H_0$ (Oe) |
|---|---|---|---|---|---|
| S1 | -46.81±1.8 | 7240±157 | 2.17±0.01 | 0.0326±0.0008 | 52.96±5.33 |
| S2 | -65.86±3.1 | 10947±403 | 2.05±0.01 | 0.0169±0.0002 | 22.16±1.89 |
| S3 | -36.03±2.3 | 8967±182 | 2.14±0.01 | 0.0245±0.0004 | 57.18±3.39 |
| S4 | -52.33±2.1 | 8987±164 | 2.10±0.01 | 0.0243±0.0004 | 22.03±3.48 |

A non-linear behaviour is also observed for sample S1 (Fig. 2(b)) above 12 GHz frequency which may be due to the other extrinsic mechanism like two magnon scattering. It is also corroborated by the enhancement of in-homogeneous broadening of sample S1 ($\Delta H_0$ =52.96 ±5.33 Oe) in comparison to sample S4 ($\Delta H_0$= 22.03 ±3.48 Oe), which indicates the interface effect. The value of α in sample S1 increases by approximately 34 % relative to S4 sample. It is a cumulative effect of spin pumping, d-d electrons hybridization and two magnon scattering [12,14,20]. Since we kept the thickness constant for the Co layer and other NM layers, we may assume that d-d electrons hybridization and two-magnon scattering will be similar contribution in all layers. Therefore, the spin-pumping enhancement at the interface of Pt/Ta is the main cause for the higher α. In addition to this, it is noted that the capping layer also affects the value of $H_K$. By changing the capping layer Pt/Ta (S1) to Au/Ta (S4), the value of $H_K$ increases by ~ 6 Oe. We cannot also rule out the effect of the phases of Co (hcp and fcc) as shown by Tokac *et al.* [21].

Now, if we compare the sample S1 to S3 (same capping layer but different seed layer), we observed lower α value and higher value of $4\pi M_{eff}$ in sample S3. However α and $4\pi M_{eff}$ values are similar for S3 and S4. Therefore, when Au is seed and/or capping layer, the Co/Au interface is significantly affecting the damping mechanism and hence spin pumping. It should be noted that both Pt and Au are heavy metals with comparable strength of spin orbit coupling [27]. However, when Pt is seed and/or capping layer the spin pumping efficiency increases e.g. in sample S1. We also noticed that $4\pi M_{eff}$ value increases when Ta/Au is taken as any of the seed or capping layer which indicates that Sample S2 and S3 have improved structural quality of the film. This is further corroborated by the reduction of *α* for sample S2. Therefore, it is concluded that the seed and capping layers are playing important role in the spin pumping for dissipating spin current into the NM layer. In order to understand the role of the combination of the two layers Ta/Pt or Ta/Au, we also fabricated samples in tri-layer structures M/Co(3)/M, where M stands for Ta, Cu and Pt for samples S5, S6 and S7, respectively.

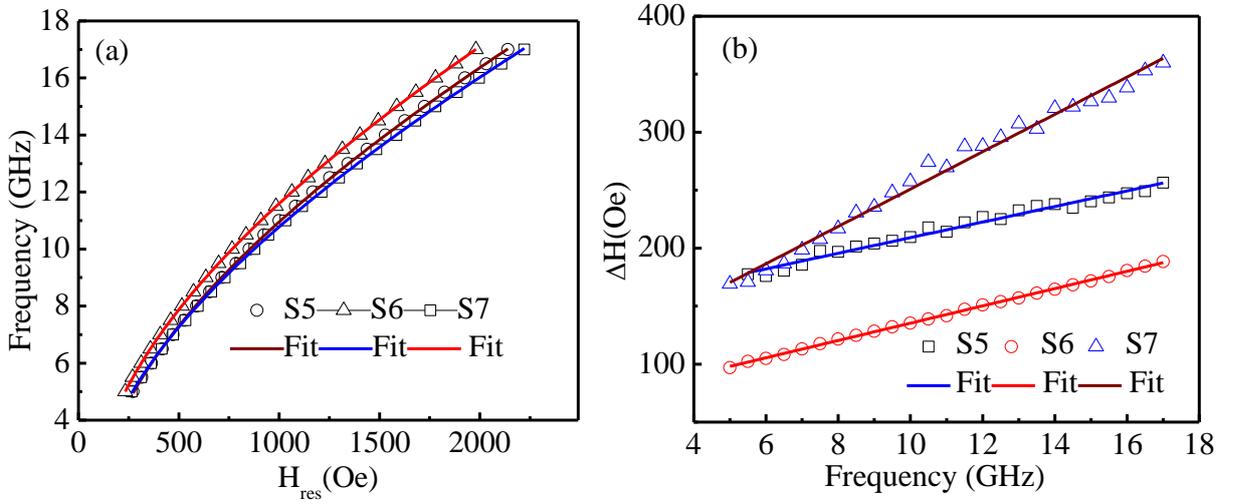

FIG. 3(a) Frequency versus resonance magnetic field ($H_{res}$) graph for sample S5, S6, S7. (b) It shows line width ($\Delta H$) versus frequency for sample S5, S6, S7. Open symbols represents the experimental data and solid lines are fit with equation (2) and (4).

Figures 3(a) and 3(b) show the frequency dependent data of $H_{res}$ and $\Delta H$, respectively. The open symbols show the experimental data and solid lines indicate the fitting with equation (2) and (4) in Figs. 3(a) and 3(b), respectively. The fitted parameters are listed in the Table III. It can be seen from the Table III that the value of $\alpha$ for S5 and S6 is nearly equal to the bulk value of Co i.e. 0.011 [26]. It is noted that Cu has low spin orbit coupling having large spin diffusion length. Therefore, the spin flipping is very weak which causes backflow of accumulated spin towards spin pumping interface. This spin back flow cancels the spin current which makes Cu as a bad spin-sink. However, Ta is a high spin orbit coupling material and hence good spin sink, but it shows low α (sample S5). This low value of α may be due to its growth as a mixed crystalline phase and the low effective spin mixing conductance at the interface. In addition to this, the value of $4\pi M_{eff}$ is higher in sample S6 while it is lower in S5. The higher value of $4\pi M_{eff}$ in sample S6 may be due to relatively better crystalline quality as Cu and Co have good lattice matching. It is known that ultra-thin films of Ta may grow in mixed phases (bcc α- phase and tetragonal β -phase) [28–30]. Therefore, lower value of $4\pi M_{eff}$ in S5 may be due to mixed phase growth of Ta, which may lead to poor structural growth of Co. In addition, it has relatively poor spin scattering efficiency due to relatively long spin diffusion length of 10 nm [31] as compared to Pt (3.5 to 10 nm) [19,20]. This is corroborated by the increase in the value of the inhomogeneous broadening $\Delta H_0$. We observe that when both seed and capping layer is Pt (S7), the value of α is high in comparison to the samples having Ta (S5) and Cu (S6). This shows the presence of large effective mixed conductance at the interface due to high spin orbit coupling of Pt. However, if we compare samples S7 with S1 which having symmetric multilayer Ta/Pt layers at top and bottom interfaces, the value of α is still higher in S1. This may be explained considering the growth properties of the Pt is affected by the Ta, which helps to induce (111) orientation of Pt [32].

Table III. The values of the fitted parameters extracted from the fitting of experimental data of $H_{res}$ and $\Delta H$ sample S5, S6, and S7 using equation (2) and (4).

| Sample | $H_K$ (Oe) | $4\pi M_{eff}$ (Oe) | g factor | α | $\Delta H_0$ (Oe) |
|---|---|---|---|---|---|
| S5 | -71.46±3.0 | 12477±7 | 2.22±0.05 | 0.0104±0.0003 | 141.8±2.37 |
| S6 | -46.25±0.5 | 15268±20 | 2.11±0.01 | 0.0109±0.0006 | 60.7±0.47 |
| S7 | -53.99±2.6 | 12548±7 | 2.15±0.04 | 0.0243±0.0005 | 89.91±4.33 |

Spin orbit coupling is an important parameter for generation of large spin current. The later is also affected by coupling between the spin, orbital angular moment, and crystal field in the system. In this context Lande-$g$ factor was evaluated from the gyromagnetic ratio that obtained by the fitting frequency dependent $H_{res}$ using equation (2). The summary of the $g$-factor and the $4\pi M_{eff}$ are listed in Table II and III, respectively, for sample S1 to S7. The highest value of $g$-factor ~ 2.22±0.05 is obtained for sample S5 which is higher than bulk value of Co (2.18) [33]. The $g$-factor is given by the combination of orbital ($\mu_L$) and spin ($\mu_S$) angular magnetic moments by the relation $\frac{g}{2} = \frac{\mu_L}{\mu_S} + 1$. Normally, in the bulk, the $\mu_L$ is absorbed by the crystal field of the material, which is protected by crystal symmetry, but, it may not be fully absorbed in case of the broken crystal symmetry at the interface. It is interesting to note that $g$-factor and $4\pi M_{eff}$ values show opposite behavior. It means for a particular sample when the $g$-factor value is higher, the value of $4\pi M_{eff}$ is lower and vice versa.

Table IV. Effective spin conductance values for sample S1 to S7.

| Sample | Sample details (In parenthesis thickness is in nm) | Effective spin mixing conductance ($g^{\uparrow\downarrow}$) ($\times 10^{18}$ m$^{-2}$) |
|---|---|---|
| S1 | Si/Ta(3)/Pt(3)/Co(3)/Pt(3)/Ta(3) | 1.86±0.08 |
| S2 | Si/Ta(3)/Au(3)/Co(3)/Au(3)/Ta(3) | 0.81±0.04 |
| S3 | Si/Ta(3)/Au(3)/Co(3)/Pt(3)/Ta(3) | 1.46±0.05 |
| S4 | Si/Ta(3)/Pt(3)/Co(3)/Au(3)/Ta(3) | 1.47±0.05 |
| S5 | Si/Ta(3)/Co(3)/Ta(3) | 0.15±0.01 |
| S6 | Si/Cu(3)/Co(3)/Cu(3) | 0.02±0.001 |
| S7 | Si/Pt(3)/Co(3)/Pt(3) | 2.00±0.08 |

To quantify the spin pumping effect, we estimated the total effective spin mixing conductance ($g^{\uparrow\downarrow}$) for all samples. The value of $g^{\uparrow\downarrow}$ can be calculated by the below expression

$$\Delta\alpha = \alpha - \alpha_0 = \frac{g\mu_B}{4\pi M_{eff} t_{FM}} g^{\uparrow\downarrow} \quad\quad\quad\quad\quad\quad\quad\quad\quad\quad\quad\quad\quad\quad\quad\quad\quad\quad\quad\quad\quad\quad\quad\quad\quad\quad\quad\quad\quad\quad\quad(5)$$

where $g$ is the Lande $g$ factor, $\mu_B$ is Bohr magneton, $t_{FM}$ is the thickness of the FM layer, and we consider the $\alpha_0$ is the bulk value of Co for evaluation of $g^{\uparrow\downarrow}$. The values are listed in Table IV for sample S1 to S7. It is known that $g^{\uparrow\downarrow}$ describes the total spin current and it gets dissipated from the FM film through its interface with NM layer by considering the backflow of the spin current. The estimated values of $g^{\uparrow\downarrow}$ are comparable to the other recent reported values [21]. It can be seen from Table IV that the value of $g^{\uparrow\downarrow}$ is highest ($2.00 \pm 0.08 \times 10^{18}$ m$^{-2}$) for sample S7 in which Pt is used as both seed and capping layers. Sample S1 exhibits a little lower $g^{\uparrow\downarrow} \sim 1.86 \pm 0.08 \times 10^{18}$ m$^{-2}$, in which Ta/Pt work as seed and capping layers. This may be due to mixed phase growth of the Ta at such thin layer, which may not work as good spin-sink layer as compared to only Pt.

To investigate the in-plane anisotropy we measured the FMR spectra with the variation of azimuthal angle at the step of 10°. Figures 4(a) and 4(b) show the azimuthal dependence of $H_{res}$ and $\Delta H$, respectively, for the samples S1 to S4. It can be observed from Figs. 4(a) and 4(b) that all samples show the uniaxial anisotropy. However, the relative strength of the anisotropy (difference between two extreme values of $H_{res}$) is higher for sample S2 having Ta/Au as seed and capping layers. In contrary sample S4 having Ta/Pt as a seed layer and Au/Ta as a capping layer shows lowest anisotropy. In addition, sample S1 with Ta/Pt as seed and capping layers shows lower anisotropy than sample S2. It means Ta/Pt seed layer reduces anisotropy in Co thin film. In order to understand the effect of bilayer and single seed layer, we measured the angular dependence of $H_{res}$ and $\Delta H$ for tri-layer samples as shown in Figs. 4(c) and 4(d), respectively. It is visible that tri-layer sample S5 having Ta as seed and capping layers has highest anisotropy among all the samples. Therefore, it is concluded that the anisotropy of Co thin film is decreased when Ta/Pt is used as seed or capping layer. However, we obtained higher spin pumping effect with Pt and Ta/Pt as seed and capping layers. Therefore, not only spin pumping damping properties, but magnetic anisotropy is also significantly affected by the seed and capping layers.

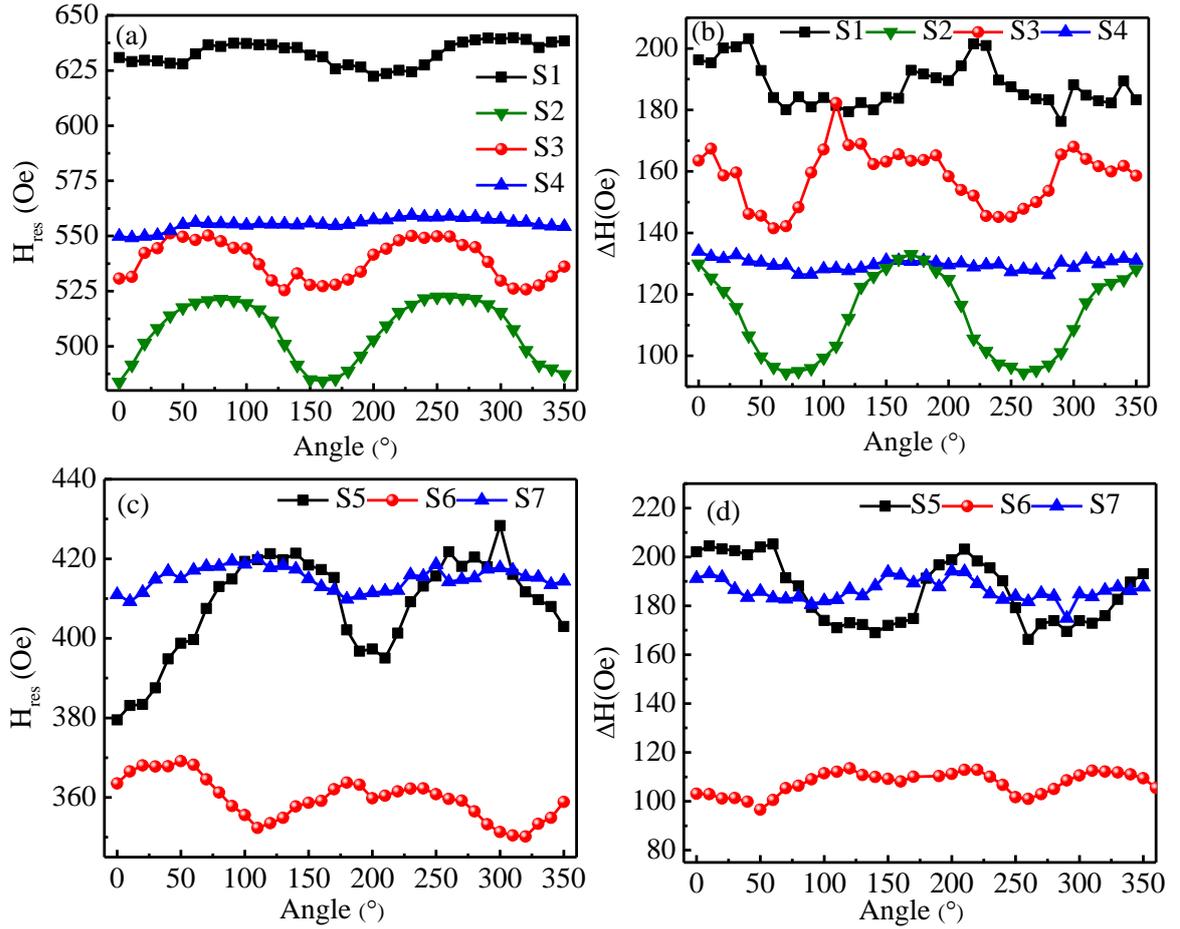

FIG. 4 Angular dependence of the $H_{res}$ of (a) sample S1 to S4 and (c) sample S5 to S7. Angular dependence of $\Delta H$ of (b) sample S1 to S4 and (d) sample S5 to S7.

## CONCLUSIONS

We presented a detailed study of magnetic dynamics of 3 nm Co thin film using FMR. The damping properties are investigated by analysing the linewidth vis-à-vis seed and capping layers in the multilayer structure. We observed higher value of damping constant i.e. $\alpha = 0.0326 \pm 0.0008$ for the symmetric multilayers having Ta/Pt as both seed and capping layers in comparison to its bulk value (0.011). This enhancement is governed by spin pumping. By considering bulk value as reference layer, we evaluated effective spin mixing conductance ~ $2.00 \pm 0.08 \times 10^{18}$ m$^{-2}$ for Pt/Co/Pt layer which is larger than *Ta/Pt/Co/Pt/Ta.* These values are in agreement with other reports [21]. It is also observed that the *g*-factor decreases with increasing effective demagnetization field with respect to seed and capping layers.

## ACKNOWLEDGEMENTS

We acknowledge National Institute of Science Education and Research, Bhubaneswar, India, DAE and DST- Nanomission (SR/NM/NS1088/2011(G)) of the Govt. of India for financial support.

**FIGURE CAPTIONS**

**Figure 1**

FIG 1(a) and (b) shows the schematics of generation of spin current in the multilayer and tri-layer sample structures (c) FMR measurement setup in which rf generator creates GHz frequency which works as a perturbation field ($h_{rf}$) perpendicular to external applied field ($H_{ext}$) and a diode detector through AC magnetic field modulation detects transmitted signal via lock-in based technique.

**Figure 2**

FIG 2(a) Resonance frequencies versus resonance magnetic field ($H_{res}$) for sample S1 to S4. Open symbols represent the experimental values while solid lines are fitted with equation (2). (b) Line width ($\Delta H$) versus resonance frequencies for sample S1 to S4. Open symbols are experimental data and solid lines are fitted with equation (4).

**Figure 3**

FIG. 3(a) Frequency versus resonance magnetic field ($H_{res}$) graph for sample S5, S6, S7. (b) It shows line width ($\Delta H$) versus frequency for sample S5, S6, S7. Open symbols represents the experimental data and solid lines are fit with equation (2) and (4).

**Figure 4**

FIG. 4 Angular dependence of the $H_{res}$ of (a) sample S1 to S4 and (c) sample S5 to S7. Angular dependence of $\Delta H$ of (b) sample S1 to S4 and (d) sample S5 to S7